\documentclass[preprint,aps]{revtex4}
\usepackage{graphicx}
\newcommand{\be}{\begin{equation}}
\newcommand{\ee}{\end{equation}}
\newcommand{\ba}{\begin{eqnarray}}
\newcommand{\ea}{\end{eqnarray}}
\newcommand{\p}{\partial}
\newcommand{\f}{\frac}

\begin{document}
\title
{Classical limit of quantum mechanics for damped driven oscillatory systems: Quantum-classical correspondence
 \vspace{0.0cm}}
\author{Jeong Ryeol Choi\footnote{E-mail: choiardor@hanmail.net } \vspace{0.0cm}}

\affiliation{Department of Electrophysics, Kyonggi University,
Yeongtong-gu, Suwon,
Gyeonggi-do 16227, Republic of Korea \vspace{0.0cm}}

\begin{abstract}
\indent
The investigation
of quantum-classical correspondence may lead
to gain a deeper understanding
of the classical limit of quantum theory.
We develop
a quantum formalism
 on the basis of a linear-invariant theorem,
which gives an exact quantum-classical correspondence for damped
oscillatory
systems that are perturbed by an arbitrary force.
Within our formalism, the quantum trajectory and expectation values of quantum
observables
are precisely coincide with
their classical counterparts
in the case where we remove the
global quantum
constant $\hbar$ from their quantum results.
In particular, we illustrate the correspondence of the quantum energy with
the classical one
in detail.
\vspace{0.0cm} \\
\end{abstract}

\maketitle
\ \ \  \\
{\bf 1. Introduction \vspace{0.2cm}}
\\
As is well known, classical mechanics (or Newtonian mechanics) is a special case of a more general
theory of physics, the so-called relativistic quantum mechanics in which quantum and relativistic
mechanics are merged.
The intrinsic outcome of classical mechanics as a low velocity limit of relativistic mechanics
has been rigorously tested and there is a common agreement for this consequence
in the community of theoretical physics.
On the other hand, the classical limit of quantum mechanics is a somewhat subtle problem.
Planck's $\hbar \rightarrow 0$ limit \cite{pla} and Bohr's $n \rightarrow \infty$ limit \cite{boh} are the oldest proposals
for the formulation of the classical limit of quantum theory.
However, there has been controversy
from the early epoch of quantum mechanics concerning
this limit through different ideas and
thoughts \cite{mank,aco,nen,lib,kuw,hua,kay}.
Accordingly, the mechanism on how to interlace the exact correspondence between the
quantum and the classical theories has not yet been fully
understood.
Man'ko and Man'ko argued
that the picture of extracting classical
mechanics with the simple limitation $\hbar \rightarrow 0$
does not have universal applicability \cite{mank}.
Some physicists
believe that quantum mechanics is not concerned with a single particle problem but an ensemble of
particles, and its $\hbar \rightarrow 0$ limit is not classical mechanics but classical statistical
mechanics instead (see Ref. \cite{hua} and references therein).
For more
different opinions concerning
the classical limit of quantum mechanics,
refer in particular to Refs. \cite{aco,kuw}.

The purpose of this research
is to
establish
a
theoretical
formalism concerning the classical limit of quantum mechanics for damped driven oscillatory systems,
which
reveals quantum and classical correspondence, without any approximation or assumption
except for the fundamental limitation $\hbar \rightarrow 0$.
Our theory is based on an invariant operator method \cite{le2,cp,jrs,gso}
which is generally used for mathematically treating quantum mechanical systems.
This method enables us to derive exact quantum mechanical solutions for time-varying Hamiltonian systems.
We will interpret
and discuss the physical meanings of our consequences in order to derive
an insight for the correspondence principle.
\\
{\bf 2. Invariant-based Dynamics and Quantum Solutions \vspace{0.2cm}} \\
To investigate quantum-classical correspondence, we consider a damped
driven harmonic oscillator of mass $m$ and frequency $\omega_0$,
whose Hamiltonian is given by
\be \hat{H} = e^{-\gamma t} \f{\hat{p}^2}{2m} + \f{1}{2} e^{\gamma t} m [\omega_0^2 \hat{q}^2 - 2f(t)\hat{q}],
\label{1} \ee
where $\gamma$ is a damping constant and $f(t)$ is a time-dependent driving force divided by $m$.
In the case of $f(t) =0$, this becomes the conventional Caldirola-Kanai (CK) Hamiltonian \cite{cal,kan}
which has been widely used in a
phenomenological
approach for the dissipation of the damped harmonic oscillator.

If we denote the classical solution of the system in configuration space as $Q(t)$,
it can be written in the form $Q(t) = Q_{\rm h} (t) + Q_{\rm p}(t)$ where $Q_{\rm h} (t)$ is a
homogeneous solution and $Q_{\rm p} (t)$ a particular solution.
From the basic algebra in
classical dynamics, we have \cite{flo}
\ba
Q_{\rm h} (t) &=& Q_{0} e^{-\gamma t/2} \cos (\omega t + \varphi), \label{1+1}  \\
Q_{\rm p} (t) &=& \int_0^t [f(t')/\omega] e^{-\gamma (t-t')/2} \sin [\omega(t-t')]  dt' , \label{1+2}
\ea
where $Q_{0}$ is the amplitude of the mechanical oscillation at $t=0$, $\omega$ is a modified frequency
which is $\omega =
({\omega_0^2 - \gamma^2/4})^{1/2}$, and $\varphi$ is an arbitrary phase.
The canonical classical solution in momentum space can also be represented
in a similar form: $P(t) = P_{\rm h} (t) + P_{\rm p}(t)$,
where $P_{\rm h} (t) = m \dot{Q}_{\rm h}(t)e^{\gamma t}$ and $P_{\rm p} (t) = m \dot{Q}_{\rm p}(t)e^{\gamma t}$.

In order to describe
quantum solutions of the system,
it is useful to introduce an invariant operator which is a powerful tool in
elucidating mechanical properties of dynamical systems that are
expressed by a time-dependent Hamiltonian like Eq. (\ref{1}).
A linear invariant operator
of the system can be derived by means of the
Liouville-von Neumann equation and it is given by (see Appendix A)
\be
\hat{I} = c [e^{-\gamma t/2} \hat{p}_{\rm p} + m ( \f{\gamma}{2} -i\omega )
e^{\gamma t/2}\hat{q}_{\rm p} ] e^{i\omega t}, \label{Lio}
\ee
where $\hat{p}_{\rm p} = \hat{p}-P_{\rm p}(t)$, $\hat{q}_{\rm p} = \hat{q}-Q_{\rm p}(t)$,
and $c =(2\hbar m\omega)^{-1/2} e^{i\chi}$ with a real constant phase $\chi$.
The eigenvalue equation of this operator can be expressed in the form
\be
\hat{I} | \phi \rangle = \lambda | \phi \rangle, \label{ive}
\ee
where $\lambda$ is the eigenvalue and $| \phi \rangle$ is the eigenstate.
We have represented the formulae of $\lambda$ and the eigenstate $\langle q| \phi \rangle$
in the configuration space in
Appendix A
including detailed derivation of them.

According to the Lewis-Riesenfeld theory \cite{le2,gco}, the wave function
that satisfies the Schr\"{o}dinger equation
is closely related to the eigenstate
of the invariant operator. In fact,
the wave function of the system in the coherent state is represented in terms
of $\langle q| \phi \rangle$
as \cite{le2}
\be
 \langle q| \psi \rangle = \langle q| \phi \rangle e^{i\theta(t)},  \label{4}
\ee
where $\theta(t)$ is a time-dependent phase. If we insert this equation together with Eq. (\ref{1}) into the Schr\"{o}dinger
equation, we have $\theta(t) = - \omega t/2$.
The wave function described here is necessary
for investigating quantum-classical correspondence through the
evolution of the system. \\
\\
{\bf 3. Correspondence between Quantum and Classical Trajectories \vspace{0.2cm}} \\
Let us now see whether the expectation values of the position
and the momentum operators
under this formalism agree with the corresponding classical trajectories or not.
Considering that the position operator is represented in terms of $\hat{I}$ as
(see Appendix A)
\be
\hat{q} = i\sqrt{\hbar/(2m\omega e^{\gamma t})}[\hat{I}e^{-i(\omega t +\chi)}-
\hat{I}^\dagger e^{i(\omega t +\chi)}]+Q_{\rm p}(t),  \label{po4}
\ee
and using Eq. (\ref{4}), we can easily
verify
that
\be
\langle \psi |\hat{q}|\psi \rangle = Q(t).
\ee
Hence, quantum expectation value of the position operator is exactly the same as the classical trajectory $Q(t)$.
In a similar way, the expectation value of
the canonical momentum is also derived such that
$\langle \psi |\hat{p}|\psi \rangle = m \dot{Q}(t) e^{\gamma t}$.
However, in general, the physical momentum in a damped system is not equivalent
to the canonical one. If we define the physical momentum operator
in the form
$\hat{p}_k = \hat{p} e^{-\gamma t}$ \cite{gre}, we readily have
\be
\langle \psi |\hat{p}_k |\psi \rangle = m \dot{Q}(t) (\equiv P_k(t)),
\ee
where $P_k(t)$ is the classical physical momentum.
We thus confirm that the linear invariant operator theory admits quantum expectation values of
$\hat{q}$ and $\hat{p}_k$ in a simple manner,
of which results precisely coincide with
the corresponding classical values.
We can regard this outcome as an initial step for verifying that
the invariant formalism of quantum mechanics
reconciles with the principle of quantum and classical correspondence.

The above consequence, however, does not mean that the quantum particle (oscillator)
follows the exact classical trajectory
that is uniquely defined. Quantum mechanics is basically non-local and there are numerous possible
paths allowed, within the width of a wave packet, for a quantum particle that has a definite initial condition.
It is impossible to indicate exactly which path the quantum
particle actually follows, but some paths may be more likely than others, especially those close to
the classically predicted path.
As a consequence of the Ehrenfest's theorem \cite{peh}, the trajectory of the quantum particle can be approximated to
that of the classical one only when the width of the quantum wave packet is sufficiently narrow.
Details of the Ehrenfest's theorem for a particular case of the system where the oscillator
is driven by a sinusoidal force are shown in Ref. \cite{ent}.
\\ \\
{\bf 4. Quantum Energy and Its Classical Limit \vspace{0.2cm}} \\
As pointed out by Hen and Kalev \cite{nen} and some other authors \cite{pol},
obtaining a quantum-classical correspondence from the test performed
at the level concerned expectation values is the key for achieving the genuine
correspondence.
Hence,
it is necessary to compare the expectation values of quantum observables with
their counterpart classical quantities.
\begin{figure}
\centering
\includegraphics[keepaspectratio=true]{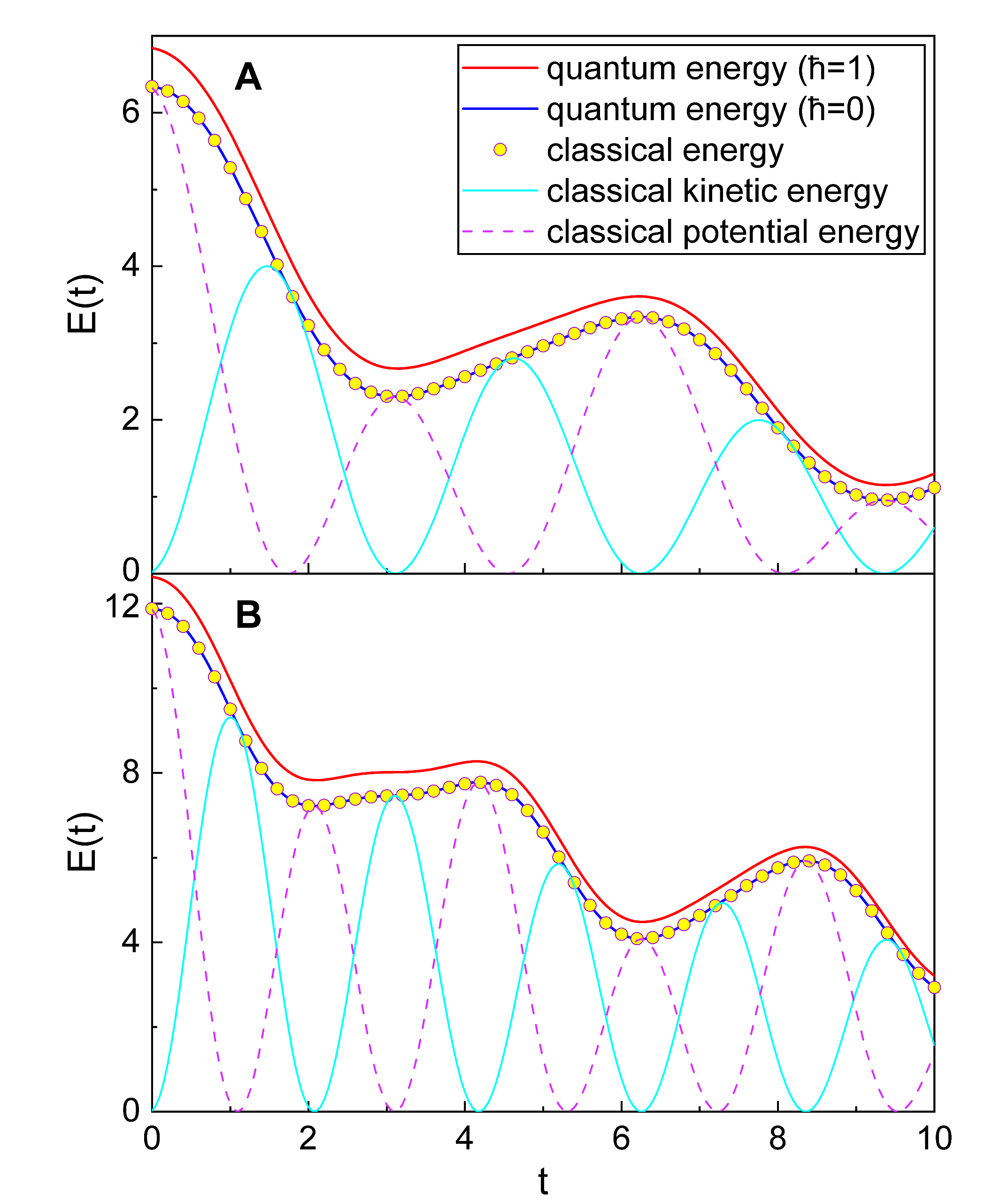}
\caption{\label{Fig1} Exact quantum energy (red line), quantum energy with $\hbar \rightarrow 0$ (blue line),
and classical mechanical energy (circle) of the oscillating cantilever in TMAFM as a function of $t$
where $k=0.5$, $a_0 = 0.3$, $D_0 = 0.5$, $\hbar = 1$, $m_{\rm eff} = 1$, $q_0 = 3$, $\gamma =0.1$, $F_{\rm ext}=0.3$,
and $\varphi =0$. The values of
$(\omega_0, \omega_{\rm d})$ are $(1, 0.3)$ for ({\bf A}) and $(1.5, 0.6)$ for ({\bf B}).
All values are taken to be dimensionless for convenience; this convention will also be used in subsequent figures.
}
\end{figure}
We now
analyze the expectation value of the quantum energy
which is one of the most common observables in the system.
Notice that quantum energy $E(t)$ for a nonconservative system is different from the expectation value of the Hamiltonian
and the expression of the energy operator, in our case, is \cite{fye2,khy}
\be
\hat{E}= e^{-2\gamma t} {\hat{p}^2}/{(2m)} + ({1}/{2}) m \omega_0^2 \hat{q}^2.  \label{E1}
\ee
After representing this operator in terms of $\hat{I}$ and $\hat{I}^\dagger$,
we are able to evaluate the expectation value of $\hat{E}$ with the help of Eq. (\ref{4}).
Through this procedure, we finally have (see Appendix B)
\be E(t) =  \f{1}{2}\hbar \Omega  + e^{-2\gamma t} \f{P^2(t)}{2m}  + \f{1}{2}  m \omega_0^2 Q^2(t), \label{6} \ee
where $\Omega = (\omega_0^2/\omega)e^{-\gamma t} $.
This is the main consequence of our present research.
The first term that contains $\hbar$ is the zero-point energy that cannot be vanished even when the displacement of the oscillator is zero.
The (quantum) energy is in general not conserved over time in dissipative systems like this,
while it is possible to predict its amount
at any given instant in time.

\begin{figure}
\centering
\includegraphics[keepaspectratio=true]{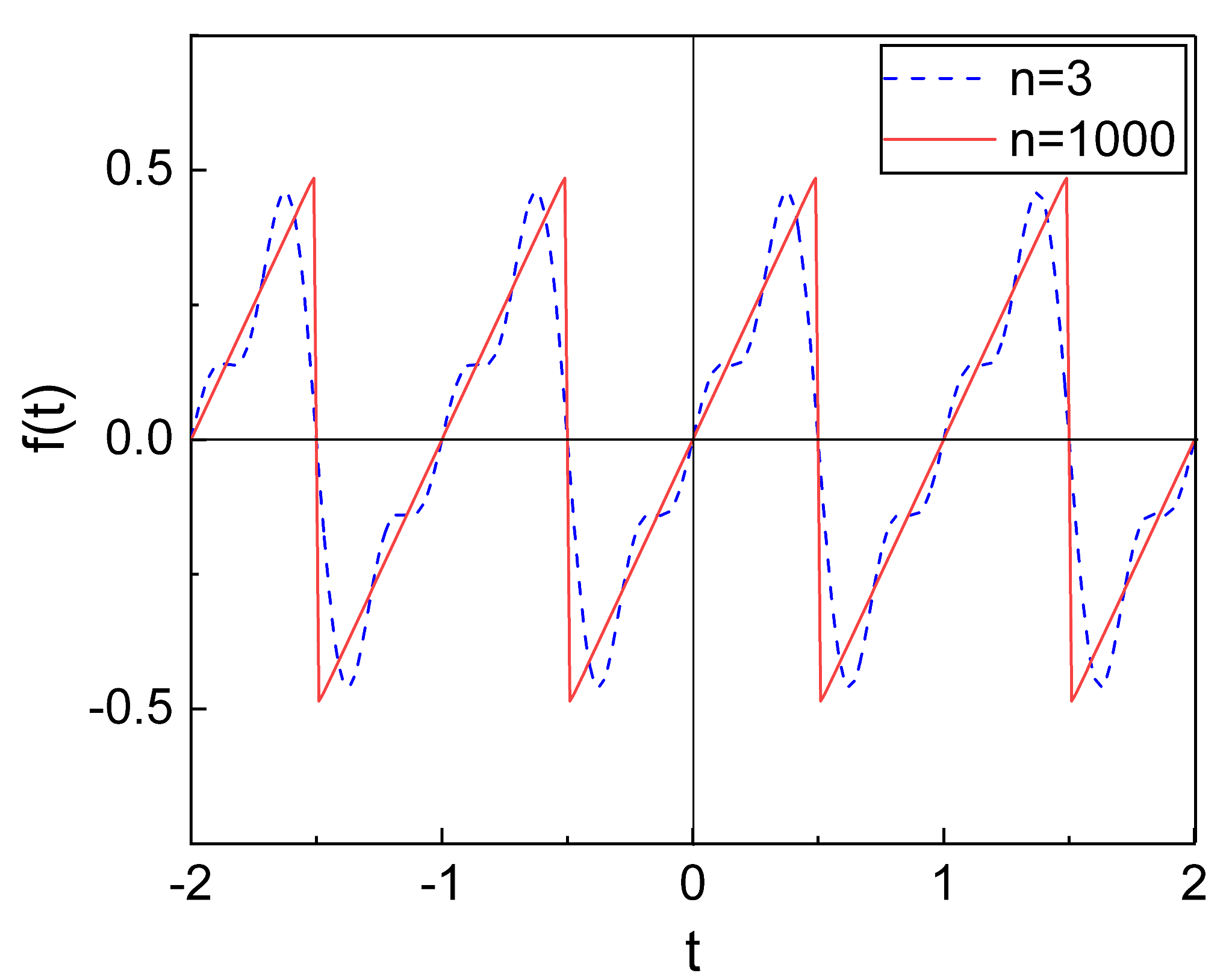}
\caption{\label{Fig2} Sawtooth driving force $f(t)$ with $f_0=1$, $m=1$, and $\tau=1$,
where the mathematical formula of $f(t)$ with a period $\tau$ is
defined in Appendix D.
All values are taken to be dimensionless for convenience.
$n$ is the natural number (see Appendix D).
We have considered $n$ up to 3 for the blue dashed line and up to 1000
for the red solid line. As $n$ increases, we can have a more exact sawtooth driving force.}
\end{figure}

For better understanding of the time behavior of Eq. (\ref{6}), let us consider
a specific system which is the
cantilever in the tapping mode atomic
force microscopy (TMAFM) \cite{spam}.
This system is widely used as a dynamic imaging technique.
For mechanical description of TMAFM, see
Appendix C.
The time evolutions of quantum energy for TMAFM are illustrated in Fig. 1
using Eq. (\ref{6}) with its comparison to the counterpart classical one.
This figure exhibits
complete consistency
between the quantum energy (with $\hbar \rightarrow 0$) and the
corresponding classical one.
We have also applied our theory to another system which
is the familiar damped harmonic oscillator driven by a periodic sawtooth
force (see Appendix D
and Fig. 2 for its mechanical description).
Sawtooth forces or signals are typically observed from
atomic force microscopy with biomolecules like proteins \cite{saw1}
and from a modulation of current density in a nuclear-fusion tokamak \cite{saw2}.
Figure 3 shows that the quantum description of this system using our theory
also coincides with the classical one.
We thus confirm that our formalism of quantum mechanics based on the linear invariant yields
exact quantum-classical correspondence.

\begin{figure}
\centering
\includegraphics[keepaspectratio=true]{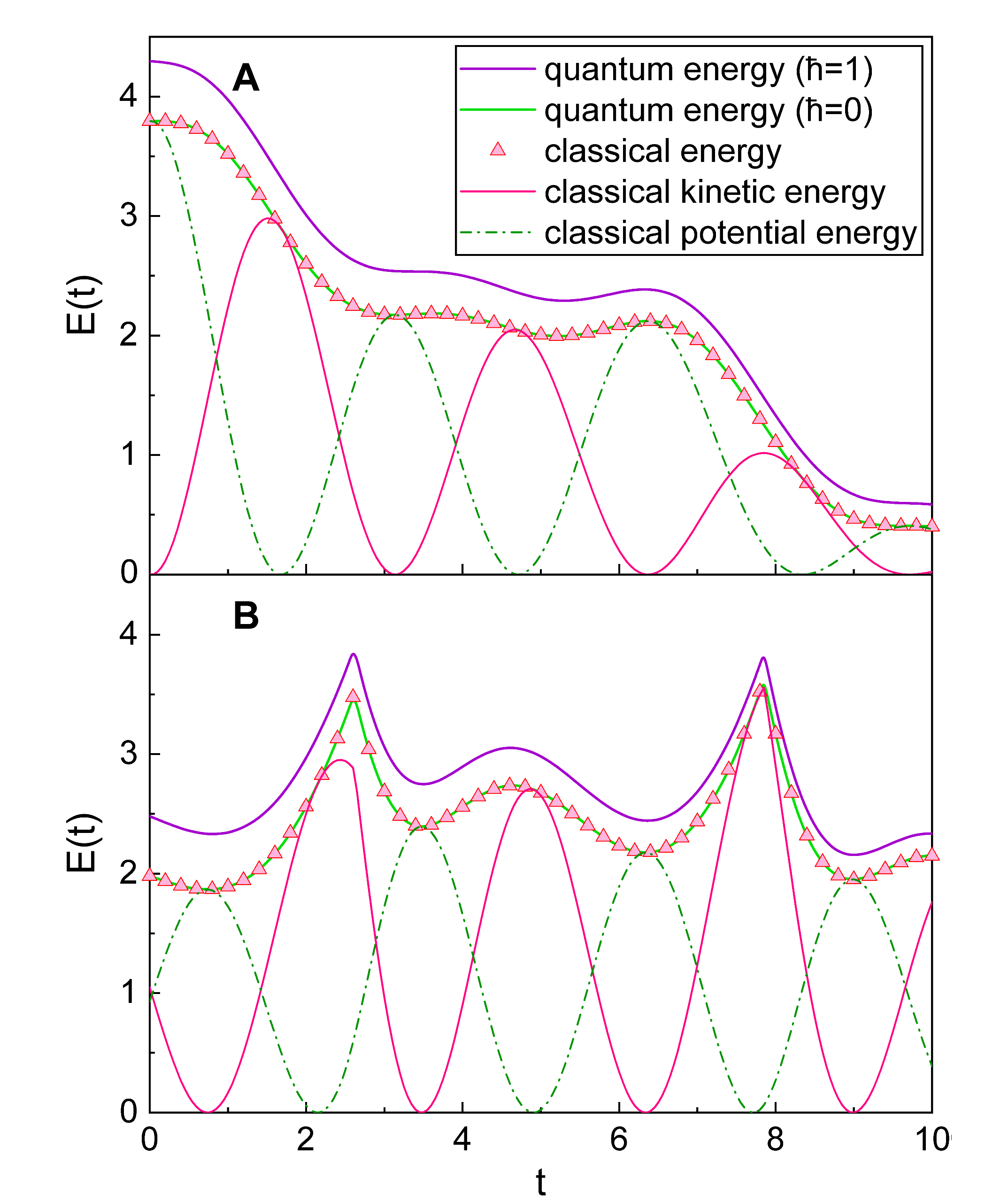}
\caption{\label{Fig3} Exact quantum energy (violet line), quantum energy with $\hbar \rightarrow 0$ (green line),
and classical mechanical energy (triangle) of the oscillator driven by the sawtooth force as a function of $t$ where
$m = 1$, $\hbar = 1$, $\gamma =0.1$, $\omega_0 = 1$, $\varphi =0$, and $n=1000$. The values of
$(q_0, \omega_{\rm d}, f_0)$ are $(3, 0.3, 1)$ for ({\bf A}) and $(1, 1.2, 2)$ for ({\bf B}).}
\end{figure}

For further analysis, let us consider the case where the driving force disappears ($f(t) \rightarrow 0$).
We can then confirm using Eq. (\ref{1+1}) that Eq. (\ref{6}) reduces to that of Ref. \cite{int3},
which is of the form
\be
E(t) =  \f{1}{2}\hbar \Omega  + E_{0} e^{-\gamma t} \left( 1+ \frac{\gamma}{2\omega_0} \cos [2({\omega} t +\varphi)-\delta]
\right), \label{7}
\ee
where $E_{0} = m\omega_0^2 Q_{0}^2/2~$  and  $\delta = \tan^{-1} ({2\omega}/{\gamma})$.
Except for the first term which is a purely quantum one, this is the well known formula of the classical mechanical energy
for the damped harmonic oscillator.
Of course, for the high displacement limit $Q_0 \gg \hbar/(m \omega)$, it is possible to neglect
the quantum effect via the use of the assumption $\hbar \sim 0$ and, consequently, the quantum energy can be successfully
approximated to the classical one.
Though the quantum energy is considered now
as a model example in order to explain the correspondence principle, one
can easily check, using the formalism developed here, that the analytical expectation values of other observables
are also in precise congruence with their classical counterparts under the limit $\hbar \rightarrow 0$.
For other formulae of quantum energies and their interpretation for this reduced system ($f(t) \rightarrow 0$),
that were derived using other methods
such as the SU(1,1) Lie algebraic approach, refer to Ref. \cite{two1}.
\\
\\
{\bf 5. Uncertainty and Correspondence Principle \vspace{0.2cm}} \\
An important feature of quantum mechanics, which distinguishes it from classical mechanics, is the appearance of
minimum uncertainty product between the arbitrary two noncommutative operators.
One cannot simultaneously know the values of position and momentum with arbitrary precision
from a quantum measurement, while
the classical theory of measurement
has nothing to do with such a limitation.

The quantum variance of an observable $\hat{\mathcal{O}}$ in the state $|\psi \rangle$ is given by
$\Delta \hat{\mathcal{O}} = [\langle \psi | \hat{\mathcal{O}}^2 | \psi \rangle
- \langle \psi | \hat{\mathcal{O}} | \psi \rangle^2]^{1/2} $.
From this identity and the use of Eq. (\ref{4}), we can straightforwardly derive the quantum uncertainty
product for position and momentum of the system and it results in
\be
\Delta \hat{q} \Delta \hat{p} =  \hbar \omega_0/(2\omega).
\ee
Because this consequence is independent of the particular solutions, $Q_{\rm p}(t)$ and $P_{\rm p}(t)$,
the driving-force
does not affect on the uncertainty product. In other words,
the uncertainty product of the system is the same as that of the un-driven damped harmonic
oscillator \cite{jrs}.
Due to the obvious inequality $\omega_0 \geq \omega$, the uncertainty principle
holds in this case.
For the case $\gamma \rightarrow 0$, this  uncertainty product reduces to $\hbar/2$ which is its minimal value allowed in
quantum mechanics for the harmonic oscillator.
On the other hand, for $\hbar \rightarrow 0$, this becomes zero, showing
the classical prediction.
\\
\\
{\bf 6. Conclusion \vspace{0.2cm}}
\\
The recent trend \cite{mey,haw} of the re-implementation of classical mechanics in particle
optics using quantum particles is a clear
testimony of the close relationship between quantum and classical mechanics.
Some essential knowledge of quantum information theory is developed on the basis of
classical-like wave properties, while the quantum nature of a physical system is unquestionable especially when
nonlocal entanglement is concerned \cite{spr}.
It may be the very common opinion that every new physical theory should not only precisely describe
facts that cannot be covered by existing theories, but must also reproduce the predictions of classical mechanics
in an appropriate classical limit.

Quantum systems exhibit various nonclassical properties such as entanglement, superposition, nonlocality,
and negative Wigner distribution function.
While such nonclassicalities are important in the next generation quantum information science,
the description of nonclassical properties is valid and reliable only when the underlying quantum formalism
used in such descriptions is precise and complete.
A formalism
of quantum theory may
be acceptable only when it gives classical results in the classical limit
($\hbar \rightarrow 0$ limit).
This is the reason why complete
quantum formalism
that obeys quantum-classical correspondence
is important.
Such a formalism
may
admit to explaining the various characteristics of dynamical systems in a reasonable and consistent way
from every possible angle.
The result for a correspondence principle that we have developed in this research
beyond simple static systems may provide
a deep insight for understanding how classical
mechanics emerges from quantum mechanics through a limiting situation.
\appendix
\section{Derivation of the Linear Invariant Operator}
From a straightforward evaluation of the Liouville-von Neumann equation,
\be
{d \hat{I}}/{d t} = {\p \hat{I}}/{\p t} +  [\hat{I},\hat{H}]/(i\hbar) = 0,
\ee
using the Hamiltonian given in
Eq. (\ref{1}) in the text, we can easily derive the linear invariant operator $\hat{I}$
that is given in Eq. (\ref{Lio}) in the text (see Ref. \cite{gso}). Notice
that the Hermitian adjoint of this operator, $\hat{I}^\dagger$, is also an invariant operator. From a combined evaluation
of the two equations for $\hat{I}$ and $\hat{I}^\dagger$, it is possible to
eliminate $\hat{p}$ and, as a consequence, the expression
for $\hat{q}$ which appeared in Eq. (\ref{po4}) in the text can be obtained.
From a similar method, we can also obtain the
expression for $\hat{p}$.
By solving the eigenvalue equation of the invariant operator, Eq. (\ref{ive}),
in the configuration
space on the basis of the technique adopted in Ref. \cite{gco},
we obtain the eigenvalue
as
\be
\lambda = \beta e^{i\omega t},
\ee
where
$\beta = -i \sqrt{{m\omega}/{(2\hbar)}} Q_{0} e^{-i(\omega t + \varphi - \chi)}$,
and the eigenstate of the form
\ba
\langle q| \phi \rangle &=& \sqrt{\f{m\omega}{\hbar\pi}} \exp \Bigg[
e^{\gamma t/2}\f{A q_{\rm p} - B q_{\rm p}^2 }{ \hbar}   + C \Bigg],  \label{(S1)}
\ea
where $q_{\rm p} = q-Q_{\rm p}(t)$ and
\ba
A &=&\sqrt{2\hbar m \omega} \beta, \\
B &=&\f{1}{2} m e^{\gamma t/2} \left(\omega + i {\gamma}/{2}\right), \\
C &=&\f{iP_{\rm p}(t)q}{\hbar}+ \f{\gamma t}{4}- \f{\beta^2}{2}- \f{|\beta|^2}{2} .
\ea
\section{Expectation Value of the Energy Operator}
We present how to evaluate the expectation value of the energy operator.
From a minor evaluation with the energy operator using the expression
of $\hat{I}$ (and its Hermitian conjugate $\hat{I}^\dagger$),
it is possible to represent the energy operator in terms of $\hat{I}$ and $\hat{I}^\dagger$ such that
\be
\hat{E} = \Bigg[\f{\hbar}{4} \left( \f{2\omega_0^2}{\omega} (2\hat{I}^\dagger \hat{I}+1)
- \varepsilon \hat{I}^2 - \varepsilon^* \hat{I}^{\dagger 2} \right)+ \sqrt{\f{\hbar}{2}}
(\Theta \hat{I} + \Theta^* \hat{I}^\dagger )\Bigg] e^{-\gamma t} + E_{\rm p} ,
\label{(S2)}
\ee
where $\varepsilon =\gamma [\gamma/(2\omega)+i]e^{-2i (\omega t + \chi)}$ and
\ba
& &\Theta = \left[ \sqrt{\f{\omega}{m}} e^{-\gamma t/2} \eta P_{\rm p}(t) + i
e^{\gamma t/2} \sqrt{\f{m}{\omega}} \omega_0^2 Q_{\rm p}(t)
\right] e^{-i(\omega t + \chi)}, \label{(S3)} \\
& &E_{\rm p} = e^{-2\gamma t} \f{P_{\rm p}^2(t)}{2m} + \f{1}{2} m
\omega_0^2 Q_{\rm p}^2(t) ,  \label{(S4)}
\ea
with $\eta =  1-i\gamma /(2\omega)$.
Now by considering the fact that the eigenvalues of $\hat{I}$ and $\hat{I}^\dagger$
are $\lambda$ and $\lambda^*$ respectively,
we can easily identify the expectation value of the energy
operator, $\langle \psi |\hat{E} |\psi \rangle$, that is given in Eq. (\ref{6}) in the text.
Notice that the $\hbar$ must not be taken simplistically to zero at
the initial stage of the evaluation under the pretext of obtaining
the classical limit. We should keep it until we arrive at the
final representation, Eq. (\ref{6}).
\section{Cantilever System}
Description of the cantilever system appears in Ref. \cite{spam}.
If we denote the effective mass of the cantilever as $m_{\rm eff}$, the force
acted on the lever is represented in the form
\be
f(t)=[F_{\rm ext} + k(D_0 - a_0 \sin \omega_{\rm d} t)]/m_{\rm eff},
\ee
where $F_{\rm ext}$ is the tip-sample force, $k(=m_{\rm eff} \omega_0^2)$ is the cantilever
spring constant, $D_0$ is the resting position of the cantilever base, $a_0$ is
the driving amplitude, and $\omega_{\rm d}$ is the drive frequency \cite{spam}.
\section{Damped Harmonic Oscillator with a Sawtooth Force}
We regard the damped harmonic oscillator to which applied an external sawtooth
force with the period
$\tau=2\pi/\omega_{\rm d}$.
The sawtooth force can be represented as $f(t)= f_0 t/(m \tau)$ for a
period $-\tau/2 < t < \tau /2$ (see Fig. 2), where $f_0$ is a constant that represents
the strength of the force. In this case, $f(t)$ can be rewritten in terms
of an infinite series such that \cite{Lc}
\be
f(t) = [{f_0}/{(\pi m)}] \sum_{n=1}^\infty \f{(-1)^{n+1}}{n} \sin (n\omega_{\rm d} t).
\label{(S5)}
\ee



\begin{references}

\bibitem{pla} M. Planck, {\it Vorlesungen uber die Theorie der W\"{a}rmestralhung} (Dover Publications, New York, 1959).
ISBN : 978-0270103878

\bibitem{boh} N. Bohr, {\it The Theory of Spectra and Atomic Constitution} (Cambridge University Press, London, 1922).
ISBN: 978-1107669819

\bibitem{lib} R. L. Liboff, On the potential $x^{2N}$ and the correspondence principle.
{\it Int. J. Theor. Phys.} {\bf 18}(3), 185-191 (1979). DOI: 10.1007/BF00670395

\bibitem{mank} O. V. Man'ko and V. I. Man'ko, Classical mechanics is not the $\hbar \rightarrow 0$
limit of quantum mechanics. {\it J. Russ. Laser Res.} {\bf 25}(5), 477-492 (2004).
DOI: 10.1023/B:JORR.0000043735.34372.8f

\bibitem{hua} X. Y. Huang, Correspondence between quantum and classical descriptions for free particles.
{\it Phys. Rev. A} {\bf 78}(2), 022109
(2008). DOI: 10.1103/PhysRevA.78.022109

\bibitem{kay} K. G. Kay, Hamiltonian formulation of quantum mechanics with semiclassical implications.
{\it Phys. Rev. A} {\bf 42}(7), 3718-3725 (1990). DOI: 10.1103/PhysRevA.46.1213

\bibitem{kuw} U. Klein, What is the limit $\hbar \rightarrow 0$ of quantum theory.
{\it Am. J. Phys.} {\bf 80}(11), 1009-1016 (2012). DOI:	10.1119/1.4751274

\bibitem{aco} A. C. Oliveira, Classical limit of quantum mechanics induced by
continuous measurements. {\it Physica A} {\bf 393}, 655-668 (2014). DOI: 10.1016/j.physa.2013.09.025

\bibitem{nen} I. Hen and A. Kalev, Classical states and their quantum correspondence.
arXiv:0701015v2 [quant-ph] (2007).

\bibitem{le2} H. R. Lewis, Jr. and W. B. Riesenfeld, {An exact quantum theory of the
time-dependent harmonic oscillator and of a charged particle in a time-dependent
electromagnetic field}. {\it J. Math. Phys.} {\bf 10}(8), 1458-1473 (1969). DOI: 10.1063/1.1664991

\bibitem{cp} M. S. Abdalla and J. R. Choi,
Propagator for the time-dependent charged oscillator via linear and quadratic invariants.
{\it Ann. Phys.} {\bf 322}(12), 2795-2810 (2007). DOI: 10.1016/j.aop.2007.01.006

\bibitem{jrs} J. R. Choi and S. Zhang,
Quantum and classical correspondence of damped-amplified oscillators.
{\it Phys. Scr.} {\bf 66}(5), 337-341 (2002). DOI: 10.1238/Physica.Regular.066a00337

\bibitem{gso} J. R. Choi and S. Ju,
Properties of the geometric phase in electromechanical oscillations of
carbon-nanotube-based nanowire resonators.
{\it Nanoscale Res. Lett.} {\bf 14}, 44 (2019). DOI: 10.1186/s11671-019-2855-8

\bibitem{cal} P. Caldirola, { Porze non conservative nella meccanica quantistica}.
 {\it Nuovo Cimento} {\bf 18}(9), 393-400 (1941). DOI: 10.1007/BF02960144

\bibitem{kan} E. Kanai,
{On the quantization of dissipative systems}.
{\it Prog. Theor. Phys.} {\bf 3}(4), 440-442 (1948). DOI: 10.1007/BF01313310

\bibitem{flo} G. Flores-Hidalgo and F. A. Barone,
{The one dimensional damped forced harmonic oscillator revisited}.
{\it Eur. J. Phys.} {\bf 32}(2), 377-379 (2011). DOI: 10.1088/0143-0807/32/2/010

\bibitem{gco} J. R. Choi and I. H. Nahm, {SU(1,1) Lie algebra applied to the general
time-dependent quadratic Hamiltonian system}.
 {\it Int. J. Theore. Phys.} {\bf 46}(1), 1-15 (2006). DOI: 10.1007/s10773-006-9050-2

\bibitem{gre} D. M. Greenberger, A critique of the major approaches to damping in quantum theory.
{\it J. Math. Phys.} {\bf 20}(5), 762-770 (1979). DOI: 10.1063/1.524148

\bibitem{peh} P. Ehrenfest, Bemerkung \"{u}ber die angen\"{a}herte g\"{u}ltigkeit der klassischen
mechanik innerhalb der quantenmechanik. {\it Z. Phys.}
{\bf 45}(7-8), 455-457 (1927). DOI: 10.1007/BF01329203

\bibitem{ent} S. Medjber, H. Bekkar, S. Menouar, and J. R. Choi,
Testing the validity of the Ehrenfest theorem beyond simple static systems: Caldirola-Kanai
oscillator driven by a time-dependent force.
{\it Chinese Phys. B} {\bf 25}(8), 080301 (2016). DOI: 10.1088/1674-1056/25/8/080301

\bibitem{pol} A. O. Bolivar, {\it Quantum-Classical Correspondence, Dynamical Quantization and the Classical
Limit} (Springer, New York, 2010). ISBN: 978-3-662-09649-9

\bibitem{fye2} R. K. Colegrave and M. S. Abdalla, A canonical description of the Fabry-Perot cavity.
{\it Optica Acta} {\bf 28}(4),
495-501 (1981). DOI: 10.1080/713820584

\bibitem{khy} C.-I. Um, K.-H. Yeon, and T. F. George, The quantum damped harmonic oscillator.
{\it Phys. Rep.} {\bf 362}(2-3), 63-192 (2002). DOI: 10.1016/S0370-1573(01)00077-1

\bibitem{spam} J. Legleiter, M. Park, B. Cusick, and T. Kowalewski, Scanning probe acceleration microscopy
(SPAM) in fluids: Mapping mechanical properties of surfaces at the nanoscale.
{\it PNAS} {\bf 103}(13), 4813-4818 (2006). DOI: 10.1073/pnas.0505628103

\bibitem{saw1} G. L. Hornyak, H. F. Tibbals, J. Dutta, and J. J. Moore,
{\it Introduction to Nanoscience \& Nanotechnology} (CRC Press, Boca Raton, 2009). ISBN: 978-1420047790

\bibitem{saw2} H. Soltwisch,
Measurement of current-density changes during sawtooth activity in a tokamak by far-infrared polarimetry.
{\it Rev. Sci. Instrum.} {\bf 59}(8), 1599-1604 (1988). DOI: 10.1063/1.1140159

\bibitem{int3} J. R. Choi,
{Interpreting quantum states of electromagnetic field in time-dependent linear media}.
{\it Phys. Rev. A} {\bf 82}(5), 055803
(2010). DOI: 10.1103/PhysRevA.82.055803

\bibitem{two1} J. R. Choi,
Analysis of quantum energy for Caldirola-Kanai Hamiltonian systems in coherent states.
{\it Results Phys.} {\bf 3}(1),
115-121 (2013). DOI: 10.1016/j.rinp.2013.06.003

\bibitem{mey} Y.-M. Wang and J.-Q. Liang,
Repulsive bound-atom pairs in an optical lattice with two-body interaction of nearest neighbors.
{\it Phys. Rev. A} {\bf 81}(4), 045601
(2010). DOI: 10.1103/PhysRevA.81.045601

\bibitem{haw} P. W. Hawkes, Examples of electrostatic electron optics:
The Farrand and Elektros microscopes and electron mirrors.
{\it Ultramicroscopy} {\bf 119}, 9-17 (2012). DOI: 10.1016/j.ultramic.2011.11.009

\bibitem{spr} M. A. Goldin, D. Francisco, and S. Ledesma, Classical images as quantum entanglement: An image
processing analogy of the GHZ experiment. {\it Opt. Commun.} {\bf 284}(7), 2089-2093 (2011).
DOI: 10.1016/j.optcom.2010.12.057

\bibitem{Lc} J. R. Choi,
Exact solution of a quantized LC circuit coupled to a power source.
{\it Phys. Scr.} {\bf 73}(6), 587-595 (2006). DOI: 10.1088/0031-8949/73/6/010

\end{references}
\end{document}